%% file: main.tex
\documentclass{article}

\ifx\HCode\UnDef\else\hypersetup{tex4ht}\fi

\usepackage[final, nonatbib]{neurips_2020_ml4ps}

\usepackage[utf8]{inputenc}

\usepackage[american]{babel}

\usepackage[usenames, dvipsnames, x11names]{xcolor}

\input{preamble/typography}

\input{preamble/crossref}

\input{preamble/biblatex}
\bibliography{main.bib}

\input{preamble/figures}

\input{preamble/tikz}

\input{preamble/commands}

\title{Physically constrained causal noise models\\ for high-contrast imaging of exoplanets}

\author{%
    Timothy D. Gebhard\\
    Max Planck Institute for Intelligent Systems, Max-Planck-Ring 4, 72076 Tübingen, Germany\\
    Institute for Particle Physics \& Astrophysics, ETH Zurich, 8092 Zurich, Switzerland\\
    \href{mailto:timothy.gebhard@tue.mpg.de}{\texttt{timothy.gebhard@tuebingen.mpg.de}}
    \And
    Markus J. Bonse\\
    Institute for Particle Physics \& Astrophysics, ETH Zurich, 8092 Zurich, Switzerland
    \And
    Sascha P. Quanz\\
    Institute for Particle Physics \& Astrophysics, ETH Zurich, 8092 Zurich, Switzerland
    \And
    Bernhard Schölkopf\\
    Max Planck Institute for Intelligent Systems, Max-Planck-Ring 4, 72076 Tübingen, Germany\\
    Department of Computer Science, ETH Zurich, 8092 Zurich, Switzerland
}

\begin{document}

    \maketitle

    \begin{abstract}
        The detection of exoplanets in high-contrast imaging (HCI) data hinges on post-processing methods to remove spurious light from the host star.
        So far, existing methods for this task hardly utilize any of the available domain knowledge about the problem explicitly.
        We propose a new approach to HCI post-processing based on a modified \emph{half-sibling regression} scheme, and show how we use this framework to combine machine learning with existing scientific domain knowledge.
        On three real data sets, we demonstrate that the resulting system performs clearly better (both visually and in terms of the SNR) than one of the currently leading algorithms.
        If further studies can confirm these results, our method could have the potential to allow significant discoveries of exoplanets both in new and archival data.
    \end{abstract}

    \input{contents/01_introduction.tex}

\input{contents/02_method.tex}

\input{contents/03_experiments.tex}

\input{contents/04_discussion.tex}

    \section*{Broader impact statement}
    The authors are not aware of any immediate ethical or societal implications of this research.
    Astrophysically, the development of new advanced post-processing algorithms for HCI data may be impactful in several ways.
    In the short and medium-term, such algorithms can be used to find new extrasolar planets in archival data or improve existing limits.
    In the long run, the architecture of these methods may also influence the design of new instruments and facilities.
    For instance, if further studies can confirm that incorporating the observing conditions into the denoising process improves the performance, this could affect decisions about which and how much additional meta-information about HCI observations is recorded.

    \begin{ack}
        The authors thank Tomas Stolker for his help in preparing the data sets.
        T.D.G. acknowledges partial funding through the Max Planck ETH Center for Learning Systems.
        Part of this work has been carried out within the framework of the NCCR PlanetS supported by the Swiss National Science Foundation.
        Finally, the authors also thank the anonymous reviewers for their helpful feedback.
        
        This research has made use of the following Python packages:
        \href{https://www.astropy.org/}{astropy}~\cite{Astropy_2013, Astropy_2018}, %
        \href{https://github.com/pydata/bottleneck}{bottleneck}, %
        \href{https://github.com/brouberol/contexttimer}{contexttimer}, %
        \href{https://github.com/nedbat/coveragepy}{coverage.py}, %
        \href{https://github.com/seperman/deepdiff}{deepdiff}, %
        \href{https://gitlab.com/pycqa/flake8}{flake8}, %
        \href{https://www.h5py.org/}{h5py}, %
        \href{https://github.com/joblib/joblib}{joblib}, %
        \href{https://jupyter.org/}{jupyter}~\cite{Jupyter_2016}, %
        \href{https://matplotlib.org/}{matplotlib}~\cite{Matplotlib_2007}, %
        \href{http://mypy-lang.org/}{mypy}, %
        \href{https://numpy.org/}{numpy}~\cite{Numpy_2020}, %
        \href{https://pandas.pydata.org/}{pandas}~\cite{Pandas_2010, Pandas_2020}, %
        \href{https://github.com/astropy/photutils}{photutils}~\cite{Photutils_2020}, %
        \href{https://pytest.org}{pytest}, %
        \href{https://scikit-learn.org/}{scikit-learn}~\cite{Sklearn_2011}, %
        \href{https://scipy.org/}{scipy}~\cite{Scipy_2020}, %
        \href{https://seaborn.pydata.org/}{seaborn}~\cite{Seaborn_2020}, %
        \href{https://github.com/tqdm/tqdm}{tqdm}%
        .
    \end{ack}

    \printbibliography

    \clearpage

    \appendix
    \input{contents/05_appendix}

\end{document}

%% file: preamble/typography.tex
\usepackage{cmap}

\usepackage[semibold]{sourcesanspro}

\usepackage[T1]{fontenc}

\usepackage{amsmath}
\usepackage{amsthm}
\usepackage{amssymb}

\usepackage{upgreek}

\usepackage{textcomp}

\usepackage{csquotes}

\usepackage{siunitx}
\usepackage[inline]{enumitem}

\usepackage{nicefrac}

\usepackage{xspace}

\usepackage{microtype}

%% file: preamble/crossref.tex
\usepackage[
    unicode=true,
    colorlinks,
    urlcolor=Blue,
    linkcolor=gray,
    citecolor=gray,
    filecolor=gray,
    pdfauthor={Timothy D. Gebhard, Markus J. Bonse, Sascha P. Quanz, Bernhard Schölkopf},
    pdftitle={Physically constrained causal noise models for high-contrast imaging of exoplanets},
    plainpages=false,
    pdfpagelabels,
]{hyperref}

\usepackage[noabbrev]{cleveref}

\usepackage{url}

\usepackage[all]{hypcap}

%% file: preamble/biblatex.tex
\usepackage[
    abbreviate=False,
    backend=biber,
    citetracker=true,
    doi=true,
    giveninits=true,
    maxbibnames=1,
    maxcitenames=1,
    natbib=true,
    sorting=none,
    style=numeric-comp,
    sortcites=true,
    url=false,
    urldate=long,
    uniquename=false,
    uniquelist=false,
]{biblatex}

\AtEveryBibitem{\clearfield{month}}
\AtEveryBibitem{\clearfield{day}}

\renewbibmacro{in:}{%
  \ifentrytype{article}{}{\printtext{\bibstring{in}\intitlepunct}}}

\DeclareFieldFormat[article]{pages}{#1}

\renewbibmacro*{volume+number+eid}{%
  \printfield{volume}%
  \printfield[parens]{number}%
  \setunit{\addcomma\space}%
  \printfield{eid}}

\DeclareFieldFormat*{eprint:arxiv}{%
  \ifhyperref
    { \href{http://arxiv.org/abs/#1}{\nolinkurl{arXiv:#1}} }
    { \nolinkurl{arXiv:#1} }
}

\usepackage{xpatch}
\xpatchbibmacro{journal+issuetitle}{%
  \setunit*{\addspace}%
  \iffieldundef{series}}
  {%
  \setunit*{\addcomma\space}%
  \iffieldundef{series}}{}{}

\renewbibmacro*{issue+date}{%
    \printfield{issue}%
}
\renewbibmacro*{author}{%
    \printnames{author}%
    \setunit*{\addspace}%
  \printtext[parens]{%
    \usebibmacro{date}}%
  \newunit}

%% file: preamble/figures.tex
\usepackage[font=small, labelfont=bf]{caption}

\usepackage{graphicx}

\usepackage{subcaption}

\usepackage{booktabs}

\usepackage[para]{threeparttable}

\usepackage{adjustbox}

\usepackage{tabularx}
\newcolumntype{C}{>{\centering\arraybackslash}X}

%% file: preamble/tikz.tex
\usepackage{standalone}

\usepackage{tikz}

\usetikzlibrary{
    decorations.pathreplacing,
    shapes, 
}

\tikzset{
    every picture/.style={font issue={\fontsize{9}{9}}},
    font issue/.style={execute at begin picture={#1\selectfont}}
}

\usepackage{dashrule}

\definecolor{Color1}{HTML}{2dbaec}  
\definecolor{Color2}{HTML}{fe7c36}  
\definecolor{Color3}{HTML}{c2dd6d}  
\definecolor{Color4}{HTML}{fa2d99}  
\definecolor{Color5}{HTML}{fedc56}  
\definecolor{Color6}{HTML}{00539c}  
\definecolor{Color7}{HTML}{984ba3}  
\definecolor{Color8}{HTML}{885448}  

\usepackage{pifont}
\newcommand{\cmark}{\ding{51}}
\newcommand{\xmark}{\ding{55}}

%% file: preamble/commands.tex
\newcommand{\eg}{e.g.,~}
\newcommand{\ie}{i.e.,~}

\newcommand{\bms}[1]{\boldsymbol{#1}}

\newcommand{\Xn}{\ensuremath{\bms{X}}\xspace}

\newcommand*\circled[1]{
    \tikz[baseline=(char.base)]{\node[shape=circle, draw, inner sep=0.5pt, font=\footnotesize] (char) {#1};}
}

%% file: contents/01_introduction.tex
\section{Introduction}
\label{sec:introduction}

\begin{figure}[t]
\minipage[b]{10cm}%
    \includestandalone[width=10cm]{figures/adi-pipeline/adi-pipeline}
    \caption{
        Finding exoplanets in HCI data requires a multi-stage post-processing pipeline.
        The most crucial step, however, is the estimation of the stellar PSF.
    }
    \label{fig:adi-pipeline}
\endminipage%
\hfill%
\minipage[b]{3.5cm}%
    \includegraphics[width=3.5cm]{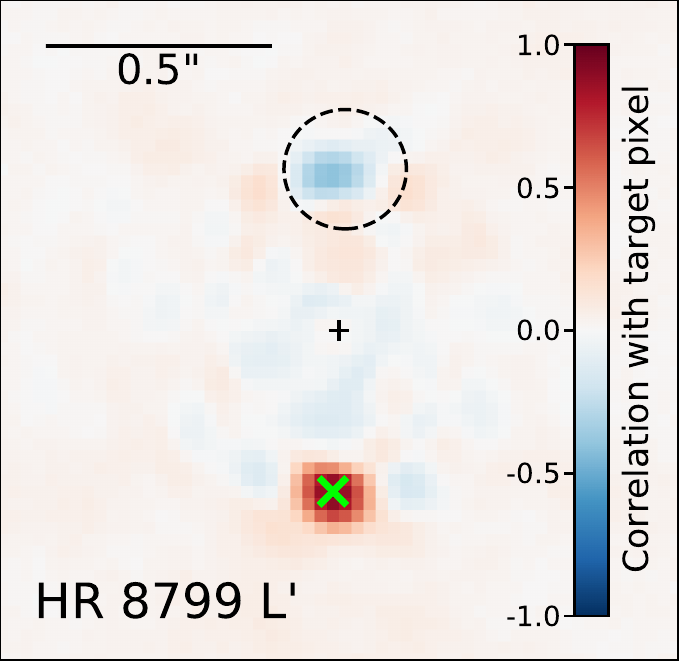}
    \caption{Example of a correlation coefficient map.}
    \label{fig:correlation-map}
\endminipage
\end{figure}

\paragraph{Context} 
High-contrast imaging (HCI) of extrasolar planets is a rapidly developing field in modern astrophysics \cite{Quanz_2010, Bowler_2016}.
Today, the detection and characterization of exoplanets through HCI is pursued at all major ground-based observatories.
While the current focus is on gas giant planets at large orbital separations, HCI at next-generation telescopes will yield the first-ever image of a terrestrial exoplanet around a nearby star \cite{Quanz_2015}. 
The most crucial step of HCI post-processing is to construct an accurate model of the stellar point spread function (PSF), which we need to subtract from the data to uncover any exoplanets close to the host star.
This is very challenging not only because the host star is several orders of magnitude brighter than any companions, but also because the PSF is non-static (\eg due to the changing atmosphere and time-variable instrument performance) and contains \emph{speckles} \cite{Bloemhof_2001}, a particular type of systematic noise that often mimics exoplanet signals.
In order to differentiate between speckles and real signals, most observations employ a technique called \emph{angular differential imaging} (ADI) \cite{Marois_2006}, where the telescope is operated in pupil-stabilized mode to record a \emph{stack} of $10^2$--$10^5$ frames over a few hours (\ie a \enquote{video} of a star and its surrounding).
Due to the Earth's rotation, the night sky (including any potential exoplanets) then appears to rotate around the target star over time.
The systematic noise, however, which emerges in the telescope, stays (approximately) fixed within the reference frame of the instrument.
We illustrate this effect together with a standard HCI/ADI post-processing pipeline in \cref{fig:adi-pipeline}.

\paragraph{Current state of the art}
In the past fifteen years, various algorithms have been proposed to estimate and remove the stellar PSF from ADI data, including LOCI~\cite{Lafreniere_2007}, ANDROMEDA~\cite{Cantalloube_2015}, LLGS~\cite{GomezGonzalez_2016}, SODINN~\cite{GomezGonzalez_2018}, FMMF~\cite{Ruffio_2017}, wavelets~\cite{Bonse_2018}, PACO~\cite{Flasseur_2018}, and TRAP~{\cite{Samland_2020}}.
In practice, one of the most popular algorithms in the community is PCA-based PSF subtraction (also known as KLIP) \cite{Soummer_2012, Amara_2012}.%
\footnote{
    Note: While the limitations of pure PCA for denoising tasks are known and possible extensions (e.g., for taking into account known properties of the noise distribution) have been proposed in other fields such as neuroscience (see, e.g., \mbox{\cite{Manjon_2015}} or \mbox{\cite{Bazin_2019}}), these innovations have not yet been adopted in the field of high-contrast imaging.
    Instead, {\enquote{vanilla}} PCA is still commonly used to detect new exoplanets; see \mbox{\cite{Bohn_2020}} for a recent example.
}
One notable weakness of many of these algorithms, and particularly of PCA/KLIP, is that they make only little to no use of the scientific domain knowledge that is available for the problem.

\paragraph{Scientific domain knowledge} 
For instance, the only real assumption that goes into \mbox{PCA/KLIP} is that the systematic noise accounts for most of the variance in the data.
We do, however, know much more about the problem.
For example:
\begin{enumerate*}[label=(\arabic*)]
    \item We know the expected spatial size of the planet signal and its temporal behavior (it moves on a circular arc with known opening angle around the star), which is determined by the known sky rotation, parametrized by the \emph{parallactic angle $\varphi(t)$}.
    \item We have a good understanding of the causal structure of the data-generating process: our data are a (potentially clipped) sum of the signal, the systematic noise (\eg speckles), as well as stochastic noise (\eg read-out noise).
    \item The theoretically expected structure of the stellar PSF has been studied extensively in the literature \cite{Bloemhof_2001, Bloemhof_2002a, Bloemhof_2002b, Bloemhof_2003, Bloemhof_2004a, Bloemhof_2004b, Bloemhof_2004c, Bloemhof_2004d, Bloemhof_2006, Bloemhof_2007, Boccaletti_2002, Sivaramakrishnan_2002, Perrin_2003, Ribak_2008}.
    One particular result is that, under certain circumstances, the speckle pattern is expected to be approximately (anti-)symmetric across the origin, meaning that if there is a speckle at position $(x, y)$, we should also see an (anti-)speckle at position $(-x, -y)$, where $(0, 0)$ is the location of the star.
    In \cref{fig:correlation-map}, we present empirical evidence for this, which we have obtained by the following experiment:
    For a given pixel $P$ at position $(x, y$) (indicated by the green cross) in the speckle-dominated regime close to the star, we compute the correlation coefficient (along the time axis) with all other spatial pixels.
    As shown in \cref{fig:correlation-map}, the region around $(-x, -y)$ (indicated by the dashed circle) is clearly anti-correlated with $P$, which can be interpreted as evidence that speckles also exhibit some degree of symmetry in practice.
    \item Additional meta-information, such as the observing conditions, are available and provide information about the temporal variation of the systematic noise.
\end{enumerate*}

\paragraph{Objectives}
Inspired by the availability (and current under-utilization) of this rich body of scientific domain knowledge, we develop a strategy that seeks to incorporate explicitly this information into a machine learning-based approach for post-processing high-contrast imaging data.

%% file: figures/adi-pipeline/adi-pipeline.tex
\begin{tikzpicture}

    \tikzset{every picture/.style={/utils/exec={\sffamily}}}
    \tikzstyle{every node}=[font=\scriptsize\sffamily, inner sep=0]

    \makeatletter
    \tikzset{
        dot diameter/.store in=\dot@diameter,
        dot diameter=0.25pt,
        dot spacing/.store in=\dot@spacing,
        dot spacing=0.44pt,
        dots/.style={
            line width=\dot@diameter,
            line cap=round,
            dash pattern=on 0pt off \dot@spacing
        }
    }
    \makeatother

    \draw [draw=none] (0.0, -0.4) rectangle (10, 3.1);

    \begin{scope}[shift={(0, 0)}]
    
        \foreach \x in {0.4, 0.3, 0.2, 0.1, 0.0} {
            \draw [draw=black, fill=white, thick] (\x, \x) rectangle ++(2.5, 2.5);
        }
        
        \draw [densely dotted, draw=black] (1.25, 1.25) -- (1.25, 2.15);
        \draw [densely dotted, draw=black] (1.25, 1.25) -- ([rotate around={-80:(1.25,1.25)}] 1.25, 2.15);
        \draw[black] ([shift=(80:0.50cm)] 1.25, 1.25) arc (80:20:0.50cm);
        \node [anchor=center, align=center] at (1.475, 1.5) {$\varphi$};
        
        \node [fill=Color5, star, star points=7, inner sep=0, minimum width=3mm] (star) at (1.25, 1.25) {};
        
        \draw [draw=none, fill=Color1] (1.25, 2.15) circle (1mm);
        \node [text=gray] at (1.25, 2.35) {\scalebox{0.7}{t\textsubscript{0}}};
        \foreach \phi [count=\n] in {20, 40, 60, 80} {
            \draw [draw=none, fill=Color1!50, rotate around={-\phi:(1.25,1.25)}] (1.25, 2.15) circle (1mm);
            \node [text=gray] at ([rotate around={-\phi:(1.25,1.25)}] 1.25, 2.35) {\scalebox{0.7}{t\textsubscript{\n}}};
        }
        \node [anchor=north west, text width=1cm, align=left] at (0.125, 2.35) {\textcolor{Color1}{\textbf{Planet}} moves on an arc};
        \draw[thick, Color1, shorten >=1.75mm, shorten <=2mm] (0.7, 2.235) edge[out=0, in=160, ->] (1.25, 2.125);
        
        \draw[thick, black, ->] ([shift=(90:0.9cm)] 1.25, 1.25) arc (90:10:0.9cm);
        
        \draw [draw=none, fill=Color3] (0.9, 1.3) circle (0.8mm);
        \draw [draw=none, fill=Color3!50, rotate around={180:(1.25,1.25)}] (0.9, 1.3) circle (0.8mm);
        \node [anchor=south west, text width=2.25cm, align=left] at (0.125, 0.125) {\textcolor{Color3}{\textbf{Speckles}} stay fixed; imperfect symmetry};
        \draw[thick, Color3!50, shorten >=1.5mm, shorten <=2mm] (0.6, 0.5) edge[out=90, in=225, ->] ([rotate around={180:(1.25,1.25)}] 0.9, 1.3);
        \draw[thick, Color3!80, shorten >=1.5mm, shorten <=2mm] (0.6, 0.5) edge[out=90, in=225, ->] (0.9, 1.3);

    \end{scope}
    \node [anchor=center, minimum height=0.5, align=center, text width=2.9cm] (PSF) at (1.45, -0.25) {input ADI stack};

    \draw [draw=none, fill=Color2!20, rounded corners] (3, 1.6) rectangle +(1.85, 1.325);
    \node [anchor=west, align=left, Color2, text width=1.75cm] at (5, 2.3) {\scriptsize\bfseries Step we want to improve!};

    \draw[thick, shorten >=1mm, shorten <=1mm, ->] (3.0, 2.4) -- ++(0.9, 0) node [midway, below=1.5mm, font=\sffamily\tiny, align=center, scale=0.85] {estimate};

    \begin{scope}[shift={(3.9, 2)}, scale=0.25]
    
        \foreach \x in {0.8, 0.6, 0.4, 0.2, 0.0} {
            \draw [draw=black, fill=white, thick] (\x, \x) rectangle ++(2.5, 2.5);
            \node [anchor=north] (PSF) at (1.65, -0.5) {PSF};
        }

        \node [fill=Color5, star, star points=7, inner sep=0, minimum width=1mm] (star) at (1.25, 1.25) {};
        
        \draw [draw=none, fill=Color3!80] (0.9, 1.3) circle (0.8mm);
        \draw [draw=none, fill=Color3!40, rotate around={180:(1.25,1.25)}] (0.9, 1.3) circle (0.8mm);

    \end{scope}

    \draw [thick, shorten >=1mm, shorten <=1mm, ->] (2.9, 0.4) -- ++(1, 0) node [midway, below=1.5mm, align=center, font=\sffamily\tiny, text width=1.1cm, scale=0.85] {subtract};
    \node [fill=black, circle, inner sep=0.5, text=white] (minus) at (3.4, 0.4) {\bfseries\raisebox{-1.25mm}{-}};

    \begin{scope}[shift={(3.9, 0)}, scale=0.25]
    
        \foreach \x in {0.8, 0.6, 0.4, 0.2, 0.0} {
            \draw [draw=black, fill=white, thick] (\x, \x) rectangle ++(2.5, 2.5);
        }
        
        \draw [draw=none, fill=Color1] (1.25, 2.15) circle (1.5mm);

        \draw[thin, black, arrows={->[scale=0.5]}] ([shift=(90:0.9cm)] 1.25, 1.25) arc (90:10:0.9cm);
        
        \node [fill=lightgray, star, star points=7, inner sep=0, minimum width=1mm] at (1.25, 1.25) {};
        
    \end{scope}
    \node [anchor=center, minimum height=0.5, align=center] at (4.3125, -0.25) {residual stack};

    \draw[thick, shorten >=1mm, shorten <=1mm] (PSF.south)+(0, -0.05) edge[out=270, in=90, ->] (minus.north);

    \draw [thick, shorten >=1mm, shorten <=1mm, ->] (4.75, 0.4) -- ++(1.0, 0) node [midway, below=1.5mm, text width=0.8cm, align=center, font=\sffamily\tiny, scale=0.85] {derotate};

    \begin{scope}[shift={(5.85, 0.125)}, scale=0.25]
        \foreach \x in {80, 60, 40, 20, 0} {
            \draw [draw=black, fill=white, thick, rotate around={\x:(1.25,1.25)}] (0, 0) rectangle ++(2.5, 2.5);
        }
        \draw [draw=none, fill=Color1] (1.25, 2.15) circle (1.5mm);
        \node [fill=lightgray, star, star points=7, inner sep=0, minimum width=1mm] at (1.25, 1.25) {};
    \end{scope}

    \draw [thick, shorten >=1mm, shorten <=1mm, ->] (6.6, 0.4) -- ++(0.90, 0) node [midway, below=1.5mm, align=center, text width=0.8cm, font=\sffamily\tiny, scale=0.85] {average};

    \draw[decorate, decoration={brace, amplitude=3pt}] (5, 0.95) -- (7.325, 0.95); 
    \node [font=\sffamily\tiny, align=center, anchor=south, text width=3.0cm, scale=0.85] at (6.15, 1.1) {Align planet signals; average out the non-systematic noise};

    \begin{scope}[shift={(7.5, 0)}]
    
        \draw [draw=black!20, thick, fill=black!10] (0, 0) rectangle ++(2.5, 2.5);
        \draw [draw=black, thick, fill=white] (1.25, 1.25) circle (1.25);
        
        \draw [draw=none, fill=Color1] (1.25, 2.15) circle (1mm);
        \node [text=Color1, yshift=-10, font=\sffamily\bfseries\tiny, text width=1.5cm, align=center] at (1.25, 2.15) {planet};
        
        \foreach \phi in {80, 60, 40, 20, 0} {
            \node [fill=lightgray, star, star points=7, inner sep=0, minimum width=3mm, rotate=\phi] (star) at (1.25, 1.25) {};
        }
        \node [text=lightgray, yshift=-7.5, font=\sffamily\bfseries\tiny, text width=1cm, align=center] at (star.south) {residual noise};
        
        \draw [draw=Color4, thin] (1.25, 2.15) circle (1.5mm);
        \draw[thick, Color4, shorten >=2.0mm, shorten <=1mm] (0.3, 2.65) edge[out=270, in=135, ->] (1.25, 2.15);

        \foreach \phi [count=\n] in {20, 40, ..., 340} {
            \draw [thin, draw=darkgray, rotate around={-\phi:(1.25,1.25)}] (1.25, 2.15) circle (1.5mm);
        }
        \draw[thick, darkgray, shorten >=2.0mm, shorten <=1mm] (2.0, 2.65) edge[out=270, in=67.5, ->] ([rotate around={-40:(1.25,1.25)}] 1.25, 2.15);
        
        \node [font=\sffamily\tiny, align=center, anchor=south, text width=3.2cm, scale=0.85] at (1.225, 2.55) {Compute SNR by comparing the \textcolor{Color4}{\textbf{signal}} with \textcolor{darkgray}{\textbf{reference apertures}}};

    \end{scope}
    \node [anchor=center, minimum height=0.5, align=center] at (8.75, -0.25) {signal estimate};
    
\end{tikzpicture}

%% file: contents/02_method.tex
\section{Method}
\label{sec:method}

\paragraph{Idea}
We propose a modified version of \emph{half-sibling regression} (HSR)~\cite{Schoelkopf_2016}, taking inspiration also from the \emph{CPM difference imaging} approach of \citet{Wang_2017}.
Our method works as follows: 
To denoise the time series of a given (spatial) pixel $Y$, we choose a set of predictor pixels $\Xn = \lbrace X_1, \ldots, X_N \rbrace$ and train (along the time axis, i.e., using different time steps as samples) a machine learning model $m$ to predict $Y$ from \Xn.
The constraint here is that \Xn must be chosen such that, under the assumption that $Y$ does contain planet signal, \Xn does \emph{not} contain planet signal.
Due to the above additivity assumption, our estimate for the planet signal in $Y$ is then given by the residual time series $\hat{Q} = Y - m(\Xn)$.
The intuition here is that, because we choose \Xn to be \emph{causally independent} of the planet signal, the prediction $m(\Xn)$ can not contain any planet signal, but only the systematic noise.
Since a planet's position is not known a priori, we have to loop over a spatial region of interest (ROI), treat every pixel in it as $Y$, and denoise its time series accordingly.
This step can easily be parallelized.
Once we have assembled the full residual stack, we obtain the final signal estimate by derotating each frame in the residual stack by its respective parallactic angle before taking the mean along the time axis (cf. \cref{fig:adi-pipeline}).

\paragraph{Choice of predictors}
For a given position $Y$, we select our predictors as illustrated in the left panel of \cref{fig:step-1} (remember that we need to loop over all possible positions $Y$ in our ROI).
Our choice is motivated by our knowledge about the temporal movement of a planet signal, as well as the expected structure of the speckle pattern.
Region~\circled{1} (in orange) is the \emph{exclusion region}, consisting of pixels which may contain signal if $Y$ \emph{at some point in time} contains a planet.
These pixels (which we can compute using our knowledge of the signal shape and the parallactic angles) must not be used as predictors, lest we run the risk of \enquote{explaining away} the signal that we are after in the first place.
The actual predictors \Xn (in green) consist of three parts:
Region~\circled{2} is chosen symmetrically opposite of $Y$ because we know from theory that if there is a speckle at $Y$, there should also be an (anti-)speckle at~\circled{2}, meaning the pixels there should be good predictors for the systematic noise.
Region~\circled{3} is chosen to capture any \enquote{local} effects around $Y$ due to the instrument, and the annulus~\circled{4} is used because we know that the systematic noise significantly depends on the radial variable.
This specific selection of predictors works well, but other choices are still part of our ongoing research.

Besides the ADI data itself, we also have access to meta-information about the observing conditions, such as wind speed or atmospheric turbulence.
These quantities are guaranteed to be causally independent from the true planet signal affecting the pixel measurements, but may contain information about the systematic noise.
The HSR framework allows us to include these data in the form of additional predictors, which, to the best of our knowledge, is something no other approach has explored so far.

\paragraph{Learning models}
Due to the flexiblity of the HSR framework, we can use virtually \emph{any} type of regression model to learn $m$.
For simplicity, we choose ridge regression with generalized cross-validation (\texttt{RidgeCV} in \texttt{sklearn}).
We can now learn such a model $m$ using the full time series for \Xn and $Y$, and call this our \emph{default HSR model}.
However, if there is a strong planet signal in $Y$ (which our predictors cannot explain as we have chosen them to be causally independent of the signal), the fit can be poor.
Therefore, we employ the following \emph{signal masking} approach consisting of two steps.

\begin{figure}[t]
\minipage[b]{6.75cm}%
    \includestandalone[width=6.75cm]{figures/step-1/step-1}
    \caption{
        In step 1, we select the predictors \Xn for a target $Y$, train the HSR models, and store candidates.
    }
    \label{fig:step-1}
\endminipage%
\hfill%
\minipage[b]{6.75cm}%
    \includestandalone[width=6.75cm]{figures/step-2/step-2}
    \caption{
        In step 2, we run a consistency test for every candidate $(Y, T)$ and compute its \emph{match fraction}.
    }
    \label{fig:step-2}
\endminipage
\end{figure}

In step 1, illustrated in \cref{fig:step-1}, we first define a grid of possible planet positions in time.
Our domain knowledge allows us to compute the \emph{expected signal form} for a given target pixel $Y$ and a time $T$ on the grid.
For every such tuple $(Y, T)$, the shape of the expected signal implies a temporal interval (where the planet signal is non-zero) that we mask out when training the model $m_{(Y, T)}$.
Once trained, we apply $m_{(Y, T)}$ to the full predictor time series and use the model's prediction to compute the residual time series $\hat{Q}_{(Y, T)} = Y - m_{(Y, T)}\left( \Xn_{(Y, T)} \right)$. 
Note that we write $\Xn_{(Y, T)}$ for the predictors of the model $m_{(Y, T)}$ because the exclusion region (and thus the predictors) depends both on the spatial position $Y$ and the assumed time $T$ at which the signal reaches its peak in $Y$.
Next, we use a simple heuristic to check if $\hat{Q}_{(Y, T)}$ contains a \enquote{signal bump} at $T$ that matches the expected planet signal shape.
If this is the case, we store $(Y, T)$ as a \emph{candidate}.
Finally, we prune our list of candidates and only keep the best (\ie highest bump) candidate for each $Y$.

In step 2, illustrated in \cref{fig:step-2}, we perform a \emph{consistency test} on all candidates from step~1.
Each candidate $(Y, T)$ implies a hypothesis about the planet's exact trajectory (\ie which pixels in the stack will be contain planet signal, and when).
Therefore, we can select other spatial positions along this implied signal path and test if their residuals also show a signal bump at the expected time.
For each candidate, we compute the \emph{match fraction}, that is, the fraction of test positions along the implied planet path that show such a consistent behavior.
We expect that only candidates due to an actual planet signal will yield a high match fraction.

Finally, we choose a threshold for the match fraction and assemble the residual stack in the following way:
For positions $Y$ with a match fraction above the threshold, we use the residuals obtained using the signal masking-based model. 
For all other positions, we use the residuals from the default HSR model.
The signal estimate is then computed from the residual stack in the usual way (cf. \cref{fig:adi-pipeline}).
Because this last step is relatively cheap computationally, we run it multiple times for different threshold values and choose the result with the highest SNR.

%% file: figures/step-1/step-1.tex
\begin{tikzpicture}

    \tikzset{every picture/.style={/utils/exec={\sffamily}}}
    \tikzstyle{every node}=[font=\scriptsize\sffamily, inner sep=0]

    \tikzset{number/.style={
        draw=black, 
        fill=white, 
        fill opacity=1.0,
        text opacity=1.0,
        circle,
        font=\scriptsize\sffamily,
        inner sep=1pt,
    }}

    \begin{scope}[shift={(0, 0)}]

        \draw [draw=black, fill=white, thick] (0, 0) rectangle ++(2.5, 2.5);

        \node [fill=Color5, star, star points=7, inner sep=0, minimum width=3mm] (star) at (1.25, 1.25) {};

        \pgfmathsetmacro{\fieldrotation}{80};
        \coordinate (center) at (1.25, 1.25);
        \coordinate (Y) at ([rotate around={-45:(center)}] 1.25, 2.15);

        \draw [draw=none, fill=Color3] (Y) circle (0.55cm);
        \draw [draw=none, fill=Color3] ([rotate around={180:(center)}] Y) circle (0.55cm);
        \fill [fill=Color3, even odd rule] (1.25, 1.25) circle[radius=0.9cm-0.1cm] circle[radius=0.9cm+0.1cm];

        \draw[draw=white, line cap=round, line width=4.5mm] ([shift=(80:0.95cm)] center) arc (80:30:0.95cm);
        \draw[draw=Color2, line cap=round, line width=4mm] ([shift=(80:0.95cm)] center) arc (80:30:0.95cm);

        \draw [draw=none, fill=Color6] (Y)+(-0.03, -0.03) rectangle ++(0.06, 0.06);
        \node [text=Color6, anchor=south east, inner sep=0.25mm, font=\bfseries\footnotesize\sffamily] at (Y) {Y};

        \node [number] at ([rotate around={30:(center)}, shift={(0.025, 0.05)}] Y) {1};
        \node [number] at ([rotate around={180:(center)}] Y)  {2};
        \node [number] at ([shift={(-0.245, -0.245)}] Y) {3};
        \node [number] at ([rotate around={90:(center)}] Y)  {4};

        \node [text=Color3, anchor=center, font=\bfseries\small\sffamily] (X) at ([rotate around={-90:(center)}, shift={(0.2, 0.2)}] Y) {X};

    \end{scope}

    \begin{scope}[shift={(2.75, 0)}]

        \draw[decorate, thick, decoration={brace, amplitude=1mm}] (0.525, 2.15) -- (2.375, 2.15); 
        \node [align=center, anchor=south, text width=2.0cm] at (1.45, 2.25) {\scalebox{0.875}{Use for training}};
        \draw[decorate, thick, decoration={brace, amplitude=1mm}] (2.425, 2.15) -- (3.075, 2.15); 
        \node [align=center, anchor=south, text width=1cm] at (2.75, 2.25) {\scalebox{0.875}{Mask\vphantom{j}}};
        \draw[decorate, thick, decoration={brace, amplitude=1mm}] (3.125, 2.15) -- (3.975, 2.15); 
        \node [align=center, anchor=south, text width=0.6cm] at (3.55, 2.25) {\scalebox{0.875}{Train\vphantom{j}}};

        \node[inner sep=0pt, anchor=west] at (0.25, 1.95) {\includegraphics[width=3.75cm]{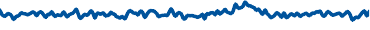}};
        \draw [draw=none, fill=white] (0.0, 1.80) rectangle ++(0.5, 0.3);
        \node[inner sep=0pt, anchor=center, font=\sffamily\bfseries\footnotesize, text=Color6] at (0.3, 1.95) {Y};
        \draw [draw=black, thick, shorten >=2mm, shorten <=1mm] (Y.east) edge[out=10, in=180, ->]  (0.3, 1.95);

        \node[inner sep=0pt, anchor=west] at (0.25, 1.6) {\includegraphics[width=3.75cm]{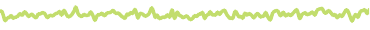}};
        \draw [draw=none, fill=white] (0.0, 1.45) rectangle ++(0.5, 0.3);
        \node[inner sep=0pt, anchor=center, text=Color3, font=\sffamily\bfseries\footnotesize] at (0.3, 1.6) {X\textsubscript{1}};
        \draw [draw=black, thick, shorten >=2mm, shorten <=1mm] (X.north) edge[out=65, in=180, ->]  (0.3, 1.6);

        \foreach \x in {0.3, 1.133, 1.766, 3.55} {
            \draw [draw=none, fill=Color3, anchor=center] (\x, 1.35) circle (0.02);
            \draw [draw=none, fill=Color3, anchor=center] (\x, 1.40) circle (0.02);
            \draw [draw=none, fill=Color3, anchor=center] (\x, 1.45) circle (0.02);
        }

        \node[inner sep=0pt, anchor=west] at (0.25, 1.20) {\includegraphics[width=3.75cm]{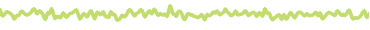}};
        \draw [draw=none, fill=white] (0.0, 1.) rectangle ++(0.5, 0.3);
        \node[inner sep=0pt, anchor=center, text=Color3, font=\sffamily\bfseries\footnotesize] at (0.3, 1.20) {X\textsubscript{N}};
        \draw [draw=black, thick, shorten >=2mm, shorten <=1mm] (X.north) edge[out=65, in=180, ->] (0.3, 1.20);
        
        \draw [draw=none, fill=white, opacity=0.7] (2.4, 1.05) rectangle ++(0.7, 1.05);
        \draw [thin, densely dotted, black] (0.50, 1.05) -- ++(0, 1.05);
        \draw [thin, densely dotted, black] (2.40, 1.05) -- ++(0, 1.05);
        \draw [thin, densely dotted, black] (3.10, 1.05) -- ++(0, 1.05);
        \draw [thin, densely dotted, black] (4.00, 1.05) -- ++(0, 1.05);

        \draw [thick, black] (2.75, 1.05) -- ++(0, 1.05);

        \draw [very thick, black] (-0.1, 0.95) -- ++(4.11, 0);


        \node[inner sep=0pt, anchor=west] at (0.25, 0.70) {\includegraphics[width=3.75cm]{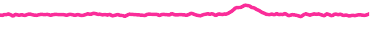}};
        \draw [draw=none, fill=white] (0.0, 0.45) rectangle ++(0.5, 0.3);
        \draw [draw=none, fill=white, opacity=0.7] (0.25, 0.45) rectangle ++(2.15, 0.3);
        \draw [draw=none, fill=white, opacity=0.7] (3.1, 0.45) rectangle ++(0.91, 0.3);
        \node[inner sep=0pt, anchor=center, text=Color4, font=\sffamily\bfseries\footnotesize] at (0.15, 0.70) {\^{Q}\textsubscript{\tiny(Y,T)}};
        
        \draw [thin, densely dotted, black] (0.50, 0.55) -- ++(0, 0.3);
        \draw [thin, densely dotted, black] (2.40, 0.55) -- ++(0, 0.3);
        \draw [thin, densely dotted, black] (3.10, 0.55) -- ++(0, 0.3);
        \draw [thin, densely dotted, black] (4.00, 0.55) -- ++(0, 0.3);

        \node [anchor=south, align=center, font=\sffamily\scriptsize, text=gray] at (0.15, 0) {Time};
        \draw [gray, ->] (0.5, 0.06) -- (4.0, 0.06);
        \foreach \x in {0.5, 0.65, ..., 3.90} {
            \draw [gray] (\x, 0.06-0.025) -- ++(0, 0.05); 
        }
 
        \draw [thick, black] (2.75, 0.55) -- ++(0, 0.3);
        \node [anchor=south west, font=\tiny\sffamily] at (2.8, 0.55) {T};
        
        \draw[decorate, thick, decoration={brace, amplitude=1mm, mirror}] (2.425, 0.5) -- (3.075, 0.5); 
        \node [align=center, anchor=north, text width=2.4cm] at (2.75, 0.365) {\scalebox{0.875}{Check for signal bump}};

    \end{scope}

\end{tikzpicture}

%% file: figures/step-2/step-2.tex
\begin{tikzpicture}

    \tikzset{every picture/.style={/utils/exec={\sffamily}}}
    \tikzstyle{every node}=[font=\scriptsize\sffamily, inner sep=0]


    \begin{scope}[shift={(0, 0)}]

        \draw [draw=black, fill=white, thick] (0, 0) rectangle ++(2.5, 2.5);

        \draw [densely dotted, draw=black] (1.25, 1.25) -- ([rotate around={80:(1.25,1.25)}] 2.1, 1.25);
        \draw [densely dotted, draw=black] (1.25, 1.25) -- ([rotate around={-40:(1.25,1.25)}] 2.1, 1.25);
        \draw[black] ([shift=(75:0.4cm)] 1.25, 1.25) arc (75:-35:0.4cm);
        \node [anchor=center, align=center] at ([rotate around={20:(1.25, 1.25)}] 1.5, 1.25) {$\varphi$};

        \draw[line width=3.5mm, line cap=round, draw=Color1] ([shift=(80:0.85cm)] 1.25, 1.25) arc (80:-40:0.85cm);

        \node [align=left, anchor=west, font=\sffamily\scriptsize, text width=1.05cm] at (0.1, 1.25) {\textcolor{Color1}{\bfseries Planet\\ trace}\\ implied by \mbox{(Y,\,T)}};
        \draw [draw=Color1, thick, shorten >=1.5mm, shorten <=1.5mm] (0.4, 1.7) edge[out=60, in=180, ->]  (1.3, 2.1);

        \node [fill=Goldenrod, star, star points=7, inner sep=0, minimum width=3mm] (star) at (1.25, 1.25) {};

        \node [draw=black, circle, thin, fill=Color4, text=white, font=\sffamily\tiny, minimum width=3mm] (Y_out) at ([rotate around={47.5:(1.25, 1.25)}] (2.1, 1.25) {Y};
        \foreach \phi [count=\n] in {75, 20, -7.5, -35} {
            \node [draw=black, circle, thin, fill=white, text=black, font=\sffamily\tiny, minimum width=3mm] (P_out_\n) at ([rotate around={\phi:(1.25, 1.25)}] (2.1, 1.25) {P\scalebox{0.75}{\textsubscript{\n}}};
        }

    \end{scope}

    \begin{scope}[shift={(2.75, 0)}]

        \pgfmathsetmacro{\panelheight}{0.4};

        \node[inner sep=0pt, anchor=west] at (0, 2.5-\panelheight*2+\panelheight/2) {\includegraphics[width=4cm]{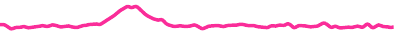}};
        \node [text=Color4, font=\sffamily\tiny, anchor=west] at (2/6*4+0.4, 2.5-2*\panelheight+0.275) {\^Q\textsubscript{\scalebox{0.7}{(Y,\,T)}}};
        \draw [] (2/6*4, 2.5-\panelheight*2) -- ++(0, \panelheight);
        \node [font=\sffamily\tiny, anchor=south west] at (2/6*4+0.05, 2.5-2*\panelheight+0.05) {T};

        \foreach \i in {1, 3, 4, 5} {
            \node[inner sep=0pt, anchor=west] at (0, 2.5-\panelheight*\i+\panelheight/2) {\includegraphics[width=4cm]{figures/step-2/expected_signal_\i}};
            
            \node[inner sep=0pt, anchor=west] at (0, 2.5-\panelheight*\i+\panelheight/2) {\includegraphics[width=4cm]{figures/step-2/observed_signal_\i}};
        }

        \foreach [count=\i] \x in {1/6, 2/6, 3/6, 4/6, 5/6} {
            \draw [] (\x*4, 2.5-\panelheight*\i) -- ++(0, \panelheight);
        }

        \foreach \i / \x in {1 / {1/6}, 4 / {4/6}, 5 / {5/6}} {
            \draw [fill=green, draw=none, opacity=0.2, anchor=west] (\x*4-0.3, 2.5-\i*\panelheight) rectangle ++(0.6, \panelheight);
            \node [text=ForestGreen, font=\sffamily\tiny] at (\x*4+0.4, 2.5-\i*\panelheight+0.275) {\cmark}; 

        }

        \draw [fill=red, draw=none, opacity=0.2, anchor=west] (3/6*4-0.3, 2.5-3*\panelheight) rectangle ++(0.6, \panelheight);
        \node [text=red, font=\sffamily\tiny] at (3/6*4+0.4, 2.5-3*\panelheight+0.275) {\xmark};

        \draw [thick, shorten >=0.5mm, shorten <=0mm] (P_out_1) edge[out=20, in=180, ->]  (0, 2.5 - 0*\panelheight - \panelheight/2);
        \draw [thick, shorten >=0.5mm, shorten <=0mm] (Y_out) edge[out=2.5, in=180, ->]  (0, 2.5 - 1*\panelheight - \panelheight/2);
        \draw [thick, shorten >=0.5mm, shorten <=0mm] (P_out_2) edge[out=-5, in=180, ->]  (0, 2.5 - 2*\panelheight - \panelheight/2);
        \draw [thick, shorten >=0.5mm, shorten <=0mm] (P_out_3) edge[out=-7.5, in=180, ->]  (0, 2.5 - 3*\panelheight - \panelheight/2);
        \draw [thick, shorten >=0.5mm, shorten <=0mm] (P_out_4) edge[out=-10, in=180, ->]  (0, 2.5 - 4*\panelheight - \panelheight/2);

        \foreach \i in {0, 1, 2, 3, 4} {
        }

        \node [align=center, anchor=center, text width=4cm] at (2, 0.2) {\textbf{Match fraction}: {\bfseries\footnotesize$\frac{\text{\textcolor{ForestGreen}{3}}}{\text{\textcolor{ForestGreen}{3}} \text{\,+\,} \text{\textcolor{red}{1}}}\text{\,=\,0.75}$}};

        \node [draw=black, fill=white, anchor=north east, thick, align=center, font=\sffamily\bfseries\tiny, inner sep=1mm] at (4-0.0, 2.5-0.0) {\textcolor{gray}{\hdashrule[0.3ex]{0.29cm}{1pt}{1pt}\,expected}\\\textcolor{Color8}{\rule[0.3ex]{0.29cm}{1pt}\,observed}};

    \end{scope}

\end{tikzpicture}

%% file: contents/03_experiments.tex
\section{Experimental evaluation and results}
\label{sec:experiments}

\paragraph{Data sets}
We showcase the performance of our improved HSR approach by applying it to three publicly available HCI data sets from the Very Large Telescope (VLT) that are known to contain exoplanets (for details, see \cref{tab:data-sets} in the appendix).
To preprocess the raw data, we use a standard pipeline based on PynPoint \cite{Amara_2012, Stolker_2019}.
Our analysis focuses on the $L'$ ($\lambda = \SI{3.80}{\micro\meter}$) and $M'$ ($\lambda = \SI{4.78}{\micro\meter}$) wavelength bands because hundreds of archival data sets are readily available and next-generation HCI instruments for the VLT (ERIS~\cite{Davies_2018}) and the ELT (METIS~\cite{Brandl_2016}) will be operating in this regime.

\paragraph{Experiments}
We apply three variants of our algorithm to our data sets: just the default HSR, HSR with signal masking (SM), and HSR with signal masking and using the observing conditions as additional predictors (SM+OC).
For a full list of all observing conditions that we used including a short description, see \cref{tab:observing-conditions} in the appendix.
As an additional pre-processing step, we median-combine blocks of frames to create data sets with an effective integration time of \SI{6.5}{\second}.
This has proven beneficial in preliminary experiments, and generally also improves the results obtained with PCA/KLIP.
For the signal masking, we use a grid with \num{50} temporal positions in step~1, and evaluate the match fraction using \num{20} test positions in step~2.
The final signal estimates are then compared to the best result obtained using PCA/KLIP, both visually and quantitatively using the signal-to-noise ratio (SNR) as defined in \citet{Mawet_2014}.

\begin{figure}
     \centering
     \begin{subfigure}[b]{32mm}
         \centering
         \includegraphics[width=\linewidth]{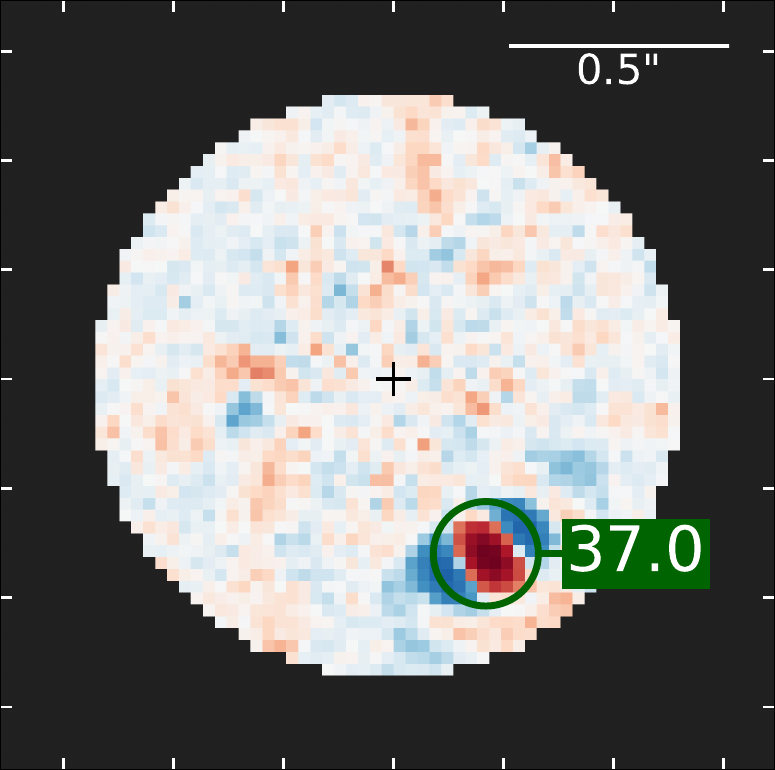}
         \caption{PCA/KLIP (baseline)}
         \label{fig:main-results-a}
     \end{subfigure}%
     \hfill%
     \begin{subfigure}[b]{32mm}
         \centering
         \includegraphics[width=\linewidth]{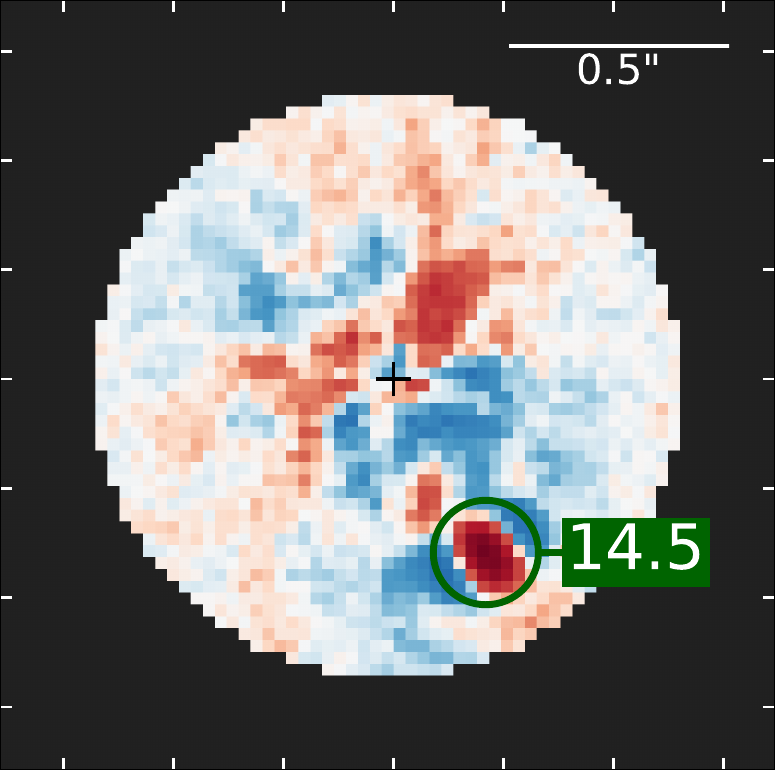}
         \caption{HSR (default)}
         \label{fig:main-results-b}
     \end{subfigure}%
     \hfill%
     \begin{subfigure}[b]{32mm}
         \centering
         \includegraphics[width=\linewidth]{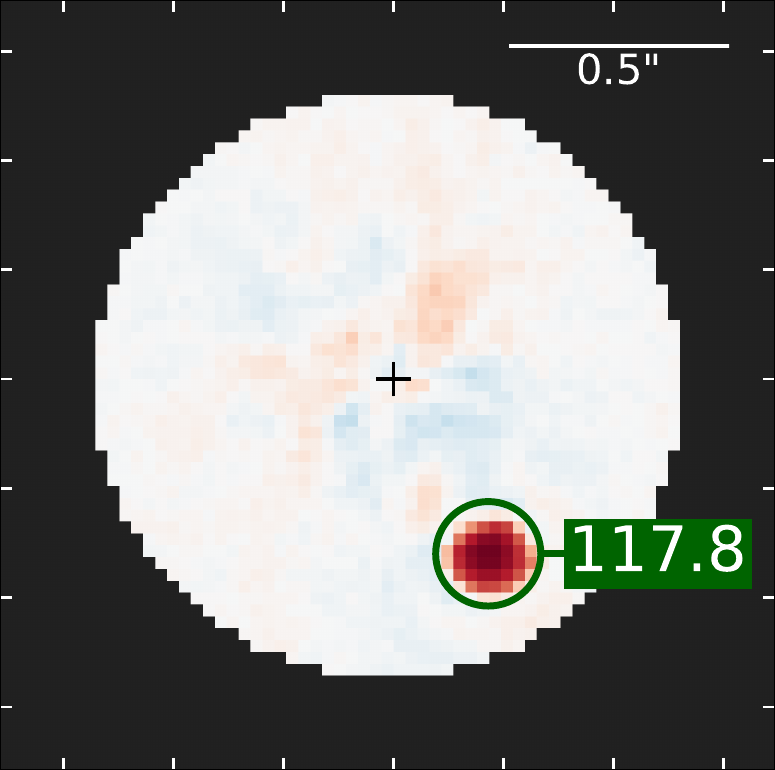}
         \caption{HSR (SM)}
         \label{fig:main-results-c}
     \end{subfigure}%
     \hfill%
     \begin{subfigure}[b]{32mm}
         \centering
         \includegraphics[width=\linewidth]{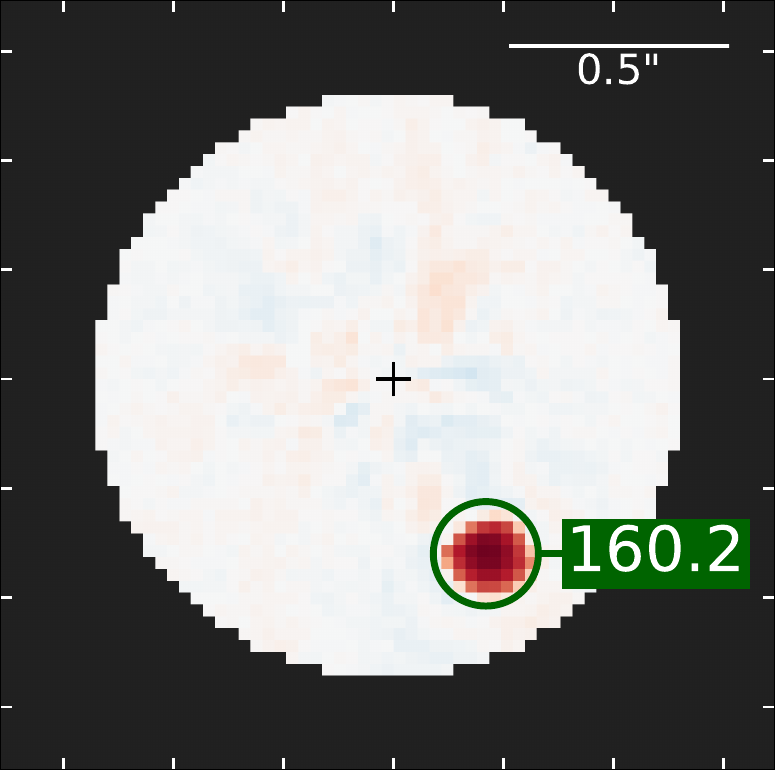}
         \caption{HSR (SM+OC)}
         \label{fig:main-results-d}
     \end{subfigure}
    \caption{
        Exemplary results (\ie signal estimates) for Beta Pictoris $L'$.
        The labels (in green) indicate the respective SNR.
        All images are oriented such that up = North.
        The color scale is a symmetric logscale from $-v_\text{max}$ to $v_\text{max}$, where $v_\text{max}$ is 1.1 times the amplitude of the  signal (\ie all plots use a different absolute scale).
    }
    \label{fig:main-results}
\end{figure}

\paragraph{Results}
We show exemplary results for the Beta Pictoris $L'$ data set in \cref{fig:main-results}.
Already visually, our proposed HSR algorithm---when used in combination with signal masking---achieves a better separation between the signal and background than the PCA baseline.
Quantitatively, we find that the achieved SNR is significantly higher for HSR than for PCA (up to a factor of 4 for Beta Pictoris $L'$).
We also notice that adding the observing conditions as additional predictors yields a substantial performance improvement.
Similar results (SNR improvements of a factor 2--3; adding OC as predictors consistently improves performance) are also found in our other two data sets; see \cref{fig:all-results} in the appendix for a full comparison.

\paragraph{Note of caution}

While these results are very encouraging, we would like to point out that the SNR alone is not yet a sufficient metric to fully characterize the performance of an HCI post-processing algorithm.
More extensive studies---for example, in the form of performance maps {\cite{JensenClem_2017}} or contrast curves---are needed to assess more thoroughly our method's ability to make new detections.

%% file: contents/04_discussion.tex
\section{Discussion and outlook}
\label{sec:discussion}

We have outlined a new algorithm for post-processing HCI data for exoplanet science and described our strategy for integrating existing scientific domain knowledge into a flexible machine learning approach. 
Our preliminary results are very encouraging and indicate that HSR could constitute a significant improvement over existing community standards.
If further studies (see above) can confirm these results, HSR could potentially enable new discoveries in hundreds of archival or new data sets.

We see the following steps for future research: 
\begin{enumerate*}[label=(\arabic*)]
\item Run additional experiments to characterize the method's performance more comprehensively (\eg compute contrast curves),
\item study more extensively different base models for the HSR, in particular non-linear models, 
\item investigate how to treat models for spatially close pixels not completely independently, 
\item study the influence of the choice of predictors, 
\item test if the signal estimate obtained with HSR can, in contrast to the PCA estimate, be used to directly measure the relative brightness of an exoplanet, and 
\item extend our method to multi-wavelength data (\ie from an integral field spectrograph), which would allow us to also incorporate additional domain knowledge about the wavelength-dependent behavior of speckles.
\end{enumerate*}

%% file: contents/05_appendix.tex
\section{Appendix}

\begin{table*}[h]
    \caption{
        Details of the three HCI data sets that we used for our experiments.
        All data were obtained with the NACO instrument at the VLT observatory and are publicly available from the ESO archive.
    }
    \label{tab:data-sets}
    \centering
    \newcommand{\esolink}[1]{\href{http://archive.eso.org/wdb/wdb/eso/sched_rep_arc/query?progid=#1}{\ttfamily#1}}
    \newcommand{\simbadlink}[1]{\href{http://simbad.u-strasbg.fr/simbad/sim-basic?Ident=#1}{#1}}
    \newcommand{\stacksize}[3]{(\num{#1}, \num{#2}, \num{#3})}
    \begin{adjustbox}{max width=\textwidth}
    \begin{threeparttable}
    \begin{tabular}{l c c l c r r r l}
    \toprule
    Target star & Filter & Date & Stack size\tnote{0} & Coronagraph & DIT (s)\tnote{1} & $\Delta \varphi$ ($^\circ$)\tnote{2} & ESO Program ID & Original reference \\
    \midrule
    \simbadlink{Beta Pictoris} & $L'$ & 2013-02-01 & \stacksize{29681}{65}{65}   & AGPM  &  0.200 & \num{83.3} & \esolink{60.A-9800(J)}  & \citet{Absil_2013} \\
    \simbadlink{Beta Pictoris} & $M'$ & 2012-11-26 & \stacksize{52122}{73}{73}   & ---   &  0.065 & \num{51.8} & \esolink{090.C-0653(D)} & \citet{Bonnefoy_2013} \\
    \simbadlink{HR 8799}       & $L'$ & 2011-09-01 & \stacksize{21043}{165}{165} & AGPM  &  0.200 & \num{32.5} & \esolink{087.C-0450(B)} & Previously unpublished.\\
    \bottomrule
    \end{tabular}
    \begin{tablenotes}
        \footnotesize
        \item[0] Format: (number of frames, frame width in pixel, frame height in pixel).
        \item[1] Detector integration time per frame
        \item[2] Field rotation of data set
    \end{tablenotes}
    \end{threeparttable}
    \end{adjustbox}
\end{table*}

\begin{table*}[h]
    \caption{
        Overview of the observing conditions used as additional predictors for the HSR model in \cref{sec:experiments}.
        The values were obtained by the Astronomical Site Monitor~(ASM) at Paranal and are directly available from the raw FITS files of the observations.
        For values were only a start and end value for each file was available we used a linear interpolation to obtain one value per frame.
    }
    \label{tab:observing-conditions}
    \centering
    \scriptsize
    \begin{tabularx}{\linewidth}{rX}
        \toprule
        Parameter name & Description \\
        \midrule
        \texttt{AIR\_MASS}                & Air mass relative to zenith (unitless). \\
        \texttt{AIR\_PRESSURE}            & Observatory ambient air pressure (in \si{\hecto\pascal}). \\
        \texttt{AVERAGE\_COHERENCE\_TIME} & Average coherence time $\tau_0$ (in \si{\second}). \\
        \texttt{M1\_TEMPERATURE}          & Superficial temperature of the primary mirror M1 (in \si{\celsius}). \\
        \texttt{OBSERVATORY\_TEMPERATURE} & Observatory ambient temperature (in \si{\celsius}). \\
        \texttt{RELATIVE\_HUMIDITY}       & Observatory ambient relative humidity (in \si{\percent}). \\
        \texttt{SEEING}                   & Observatory Seeing (in arcsec). \\
        \texttt{WIND\_SPEED}              & Observatory ambient wind speed (in \si{\meter\per\second}). \\
        \texttt{COS\_WIND\_DIRECTION}     & Cosine of the observatory ambient wind direction (unitless). \\
        \texttt{SIN\_WIND\_DIRECTION}     & Sine of the observatory ambient wind direction (unitless). \\
        \bottomrule
    \end{tabularx}
\end{table*}

\begin{figure}[h]
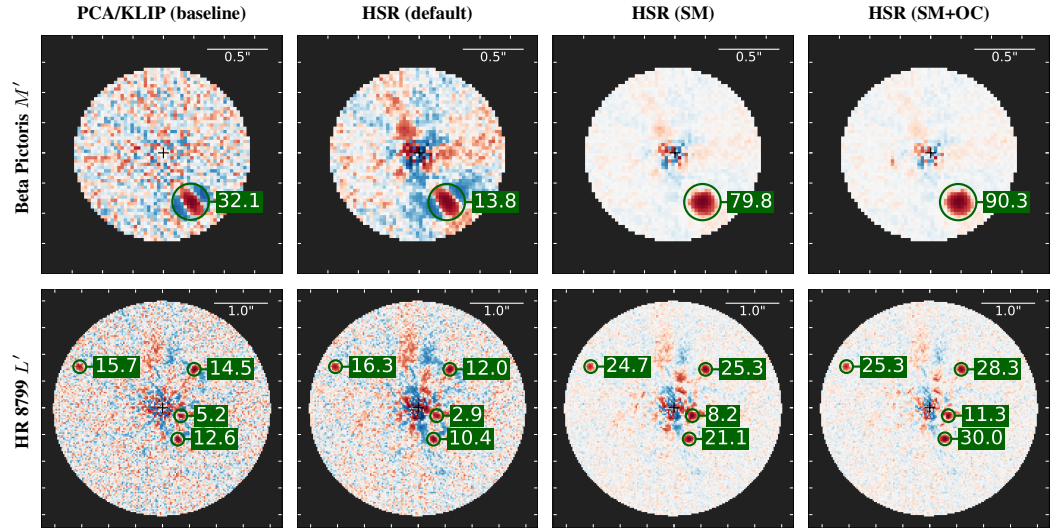

    \centering
    \setlength{\tabcolsep}{0mm}
    \newcommand{\rot}[1]{\adjustbox{margin=0.75mm}{\rotatebox[origin=c]{90}{\textbf{\scriptsize#1}}}}
    \newcommand{\bth}[1]{\adjustbox{margin=0.75mm}{\textbf{\scriptsize#1}}}
    \newcommand{\imginclude}[1]{\adjincludegraphics[valign=M, width=32mm, margin=1mm]{#1}}
    \begin{tabular}{ccccc}
        & \bth{PCA/KLIP (baseline)} & \bth{HSR (default)} & \bth{HSR (SM)} & \bth{HSR (SM+OC)}\\
        \rot{Beta Pictoris $M'$} & 
        \imginclude{figures/results/Beta_Pictoris__Mp/pca} &%
        \imginclude{figures/results/Beta_Pictoris__Mp/baseline} &%
        \imginclude{figures/results/Beta_Pictoris__Mp/sm} &%
        \imginclude{figures/results/Beta_Pictoris__Mp/sm_oc} \\%
        \rot{HR 8799 $L'$} & 
        \imginclude{figures/results/HR_8799__Lp/pca} &%
        \imginclude{figures/results/HR_8799__Lp/baseline} &%
        \imginclude{figures/results/HR_8799__Lp/sm} &%
        \imginclude{figures/results/HR_8799__Lp/sm_oc} \\%
    \end{tabular}
    \caption{
        Additional plots for the main experiment from \cref{sec:experiments} showing the results for the data \mbox{Beta~Pictoris~$M'$} and HR~8799~$L'$.
        Again, all images are oriented such that up = North, labels indicate the respective SNR, and each plot uses a symmetric logscale.
        We find our results from \cref{fig:main-results} confirmed: HSR in combination with signal masking is significantly better than PCA, both visually and in terms of SNR.
        Furthermore, adding the observing conditions as additional predictors improves the SNR even further.
    }
    \label{fig:all-results}
\end{figure}

%% file: main.bib
@Article{         Absil_2013,
  title         = {Searching for companions down to 2 {AU} from $\upbeta$
                  Pictoris using the $L'$-band {AGPM} coronagraph on
                  {VLT}/{NACO}},
  author        = {O. Absil and others},
  year          = {2013},
  month         = {11},
  journal       = {Astronomy {\&} Astrophysics},
  volume        = {559},
  pages         = {L12},
  doi           = {10.1051/0004-6361/201322748},
  eprint        = {1311.4298},
  eprinttype    = {arXiv}
}

@Article{         Amara_2012,
  title         = {{PYNPOINT}: an image processing package for finding
                  exoplanets},
  author        = {Amara, Adam and Quanz, Sascha P.},
  year          = {2012},
  journal       = {Monthly Notices of the Royal Astronomical Society},
  volume        = {427},
  number        = {2},
  pages         = {948--955},
  doi           = {10.1111/j.1365-2966.2012.21918.x},
  eprint        = {1207.6637},
  eprinttype    = {arXiv}
}

@Article{         Astropy_2013,
  title         = {Astropy: A community Python package for astronomy},
  author        = {{Astropy Collaboration} and Thomas P. Robitaille and Erik
                  J. Tollerud and others},
  year          = {2013},
  month         = {9},
  journal       = {Astronomy {\&} Astrophysics},
  volume        = {558},
  pages         = {A33},
  doi           = {10.1051/0004-6361/201322068},
  eprint        = {1307.6212},
  eprinttype    = {arXiv}
}

@Article{         Astropy_2018,
  title         = {The Astropy Project: Building an Open-science Project and
                  Status of the v2.0 Core Package},
  author        = {{Astropy Collaboration} and A. M. Price-Whelan and B. M.
                  Sipőcz and others},
  year          = 2018,
  month         = 8,
  journal       = {The Astronomical Journal},
  volume        = 156,
  number        = 3,
  pages         = 123,
  doi           = {10.3847/1538-3881/aabc4f},
  eprint        = {1801.02634},
  eprinttype    = {arXiv}
}

@Article{         Bazin_2019,
  title         = {Denoising High-Field Multi-Dimensional {MRI} With Local
                  Complex {PCA}},
  author        = {Pierre-Louis Bazin and Anneke Alkemade and Wietske van der
                  Zwaag and Matthan Caan and Martijn Mulder and Birte U.
                  Forstmann},
  year          = {2019},
  month         = oct,
  journal       = {Frontiers in Neuroscience},
  volume        = {13},
  doi           = {10.3389/fnins.2019.01066}
}

@Article{         Bloemhof_2001,
  title         = {Behavior of Remnant Speckles in an Adaptively Corrected
                  Imaging System},
  author        = {E. E. Bloemhof and R. G. Dekany and M. Troy and B. R.
                  Oppenheimer},
  year          = {2001},
  month         = {9},
  journal       = {The Astrophysical Journal},
  volume        = {558},
  number        = {1},
  pages         = {L71--L74},
  doi           = {10.1086/323494}
}

@InProceedings{   Bloemhof_2002a,
  title         = {Statistics of remnant speckles in an adaptively corrected
                  imaging system},
  author        = {Eric E. Bloemhof},
  year          = {2002},
  month         = {2},
  doi           = {10.1117/12.454811},
  booktitle     = {Adaptive Optics Systems and Technology {II}},
  publisher     = {{SPIE}},
  editor        = {Robert K. Tyson and Domenico Bonaccini and Michael C.
                  Roggemann}
}

@Article{         Bloemhof_2002b,
  title         = {Suppression of Speckle Noise by Speckle Pinning in
                  Adaptive Optics},
  author        = {E. E. Bloemhof},
  year          = {2002},
  month         = {12},
  journal       = {The Astrophysical Journal},
  volume        = {582},
  number        = {1},
  pages         = {L59--L62},
  doi           = {10.1086/346100}
}

@InProceedings{   Bloemhof_2003,
  title         = {Speckles in a highly corrected adaptive optics system},
  author        = {E. E. Bloemhof},
  year          = {2003},
  month         = {12},
  doi           = {10.1117/12.507241},
  booktitle     = {Astronomical Adaptive Optics Systems and Applications},
  publisher     = {{SPIE}},
  editor        = {Robert K. Tyson and Michael Lloyd-Hart}
}

@Article{         Bloemhof_2004a,
  title         = {Anomalous intensity of pinned speckles at high adaptive
                  correction},
  author        = {E. E. Bloemhof},
  year          = {2004},
  month         = {1},
  journal       = {Optics Letters},
  volume        = {29},
  number        = {2},
  pages         = {159},
  doi           = {10.1364/ol.29.000159}
}

@Article{         Bloemhof_2004b,
  title         = {Remnant Speckles in a Highly Corrected Coronagraph},
  author        = {E. E. Bloemhof},
  year          = {2004},
  month         = {6},
  journal       = {The Astrophysical Journal},
  volume        = {610},
  number        = {1},
  pages         = {L69--L72},
  doi           = {10.1086/423172}
}

@InProceedings{   Bloemhof_2004c,
  title         = {Speckle noise in highly corrected coronagraphs},
  author        = {Eric E. Bloemhof},
  year          = {2004},
  month         = {10},
  doi           = {10.1117/12.560370},
  booktitle     = {Advanced Wavefront Control: Methods, Devices, and
                  Applications {II}},
  publisher     = {{SPIE}},
  editor        = {John D. Gonglewski and Mark T. Gruneisen and Michael K.
                  Giles}
}

@Article{         Bloemhof_2004d,
  title         = {Static point-spread function correction dominating
                  higher-order speckle terms at high adaptive correction},
  author        = {E. E. Bloemhof},
  year          = {2004},
  month         = {10},
  journal       = {Optics Letters},
  volume        = {29},
  number        = {20},
  pages         = {2333},
  doi           = {10.1364/ol.29.002333}
}

@InProceedings{   Bloemhof_2006,
  title         = {Suppression of speckles at high adaptive correction using
                  speckle symmetry},
  author        = {E. E. Bloemhof},
  year          = {2006},
  month         = {8},
  doi           = {10.1117/12.681477},
  booktitle     = {Instruments, Methods, and Missions for Astrobiology {IX}},
  publisher     = {{SPIE}},
  editor        = {Richard B. Hoover and Gilbert V. Levin and Alexei Y.
                  Rozanov}
}

@Article{         Bloemhof_2007,
  title         = {Feasibility of symmetry-based speckle noise reduction for
                  faint companion detection},
  author        = {E. E. Bloemhof},
  year          = {2007},
  journal       = {Optics Express},
  volume        = {15},
  number        = {8},
  pages         = {4705},
  doi           = {10.1364/oe.15.004705}
}

@Article{         Boccaletti_2002,
  title         = {Speckle Symmetry with High-Contrast Coronagraphs},
  author        = {A. Boccaletti and P. Riaud and D. Rouan},
  year          = {2002},
  journal       = {Publications of the Astronomical Society of the Pacific},
  volume        = {114},
  number        = {792},
  pages         = {132--136},
  doi           = {10.1086/338914}
}

@Article{         Bohn_2020,
  title         = {Two Directly Imaged, Wide-orbit Giant Planets around the
                  Young, Solar Analog {TYC} 8998-760-1},
  author        = {Alexander J. Bohn and others},
  year          = {2020},
  month         = {7},
  journal       = {The Astrophysical Journal},
  volume        = {898},
  number        = {1},
  pages         = {L16},
  doi           = {10.3847/2041-8213/aba27e},
  eprint        = {2007.10991},
  eprinttype    = {arXiv}
}

@Article{         Bonnefoy_2013,
  title         = {The near-infrared spectral energy distribution of
                  $\upbeta$ Pictoris b},
  author        = {M. Bonnefoy and A. Boccaletti and A.-M. Lagrange and F.
                  Allard and C. Mordasini and H. Beust and G. Chauvin and J.
                  H. V. Girard and D. Homeier and D. Apai and S. Lacour and
                  D. Rouan},
  year          = {2013},
  month         = {7},
  journal       = {Astronomy {\&} Astrophysics},
  volume        = {555},
  pages         = {A107},
  doi           = {10.1051/0004-6361/201220838},
  eprint        = {1302.1160},
  eprinttype    = {arXiv}
}

@Article{         Bonse_2018,
  title         = {Wavelet based speckle suppression for exoplanet imaging
                  --- Application of a de-noising technique in the time
                  domain},
  author        = {Markus J. Bonse and Sascha P. Quanz and Adam Amara},
  year          = {2018},
  eprint        = {1804.05063},
  eprinttype    = {arXiv}
}

@Article{         Bowler_2016,
  title         = {Imaging Extrasolar Giant Planets},
  author        = {Brendan P. Bowler},
  year          = {2016},
  month         = {10},
  journal       = {Publications of the Astronomical Society of the Pacific},
  volume        = {128},
  number        = {968},
  pages         = {102001},
  doi           = {10.1088/1538-3873/128/968/102001},
  eprint        = {1605.02731},
  eprinttype    = {arXiv}
}

@InProceedings{   Brandl_2016,
  title         = {Status of the mid-infrared E-{ELT} imager and spectrograph
                  {METIS}},
  author        = {Bernhard R. Brandl and others},
  year          = {2016},
  month         = {8},
  doi           = {10.1117/12.2233974},
  booktitle     = {Ground-based and Airborne Instrumentation for Astronomy
                  {VI}},
  publisher     = {{SPIE}},
  editor        = {Christopher J. Evans and Luc Simard and Hideki Takami}
}

@Article{         Cantalloube_2015,
  title         = {Direct exoplanet detection and characterization using the
                  {ANDROMEDA} method: Performance on {VLT}/{NaCo} data},
  author        = {F. Cantalloube and D. Mouillet and L. M. Mugnier and J.
                  Milli and O. Absil and C. A. Gomez Gonzalez and G. Chauvin
                  and J.-L. Beuzit and A. Cornia},
  year          = {2015},
  journal       = {Astronomy {\&} Astrophysics},
  volume        = {582},
  pages         = {A89},
  doi           = {10.1051/0004-6361/201425571},
  eprint        = {1508.06406},
  eprinttype    = {arXiv}
}

@InProceedings{   Davies_2018,
  title         = {{ERIS}: revitalising an adaptive optics instrument for the
                  {VLT}},
  author        = {Richard Davies and others},
  year          = {2018},
  month         = {7},
  doi           = {10.1117/12.2311480},
  booktitle     = {Ground-based and Airborne Instrumentation for Astronomy
                  {VII}},
  publisher     = {{SPIE}},
  editor        = {Hideki Takami and Christopher J. Evans and Luc Simard},
  eprint        = {1807.05089},
  eprinttype    = {arXiv}
}

@Article{         Flasseur_2018,
  title         = {Exoplanet detection in angular differential imaging by
                  statistical learning of the nonstationary patch
                  covariances},
  author        = {Olivier Flasseur and Loïc Denis and Éric Thiébaut and
                  Maud Langlois},
  year          = {2018},
  journal       = {Astronomy {\&} Astrophysics},
  volume        = {618},
  pages         = {A138},
  doi           = {10.1051/0004-6361/201832745}
}

@Article{         GomezGonzalez_2016,
  title         = {Low-rank plus sparse decomposition for exoplanet detection
                  in direct-imaging {ADI} sequences},
  author        = {C. A. {Gomez~Gonzalez} and O. Absil and P.-A. Absil and M.
                  {Van~Droogenbroeck} and D. Mawet and J. Surdej},
  year          = {2016},
  journal       = {Astronomy {\&} Astrophysics},
  volume        = {589},
  pages         = {A54},
  doi           = {10.1051/0004-6361/201527387},
  eprint        = {1602.08381},
  eprinttype    = {arXiv}
}

@Article{         GomezGonzalez_2018,
  title         = {Supervised detection of exoplanets in high-contrast
                  imaging sequences},
  author        = {C. A. {Gomez~Gonzalez} and O. Absil and M.
                  {Van~Droogenbroeck}},
  year          = {2018},
  journal       = {Astronomy {\&} Astrophysics},
  volume        = {613},
  pages         = {A71},
  doi           = {10.1051/0004-6361/201731961},
  eprint        = {1712.02841},
  eprinttype    = {arXiv}
}

@Article{         JensenClem_2017,
  title         = {{A New Standard for Assessing the Performance of High
                  Contrast Imaging Systems}},
  author        = {Rebecca Jensen-Clem and Dimitri Mawet and Carlos A. Gomez
                  Gonzalez and Olivier Absil and Ruslan Belikov and Thayne
                  Currie and Matthew A. Kenworthy and Christian Marois and
                  Johan Mazoyer and Garreth Ruane and Angelle Tanner and
                  Faustine Cantalloube},
  year          = {2017},
  month         = {12},
  journal       = {The Astronomical Journal},
  volume        = {155},
  number        = {1},
  pages         = {19},
  doi           = {10.3847/1538-3881/aa97e4}
}

@Conference{      Jupyter_2016,
  title         = {Jupyter Notebooks -- a publishing format for reproducible
                  computational workflows},
  author        = {Thomas Kluyver and Benjamin Ragan-Kelley and Fernando
                  Pérez and Brian Granger and Matthias Bussonnier and
                  Jonathan Frederic and Kyle Kelley and Jessica Hamrick and
                  Jason Grout and Sylvain Corlay and Paul Ivanov and
                  Dami{'a}n Avila and Safia Abdalla and Carol Willing},
  year          = {2016},
  pages         = {87--90},
  booktitle     = {Positioning and Power in Academic Publishing: Players,
                  Agents and Agendas},
  editor        = {F. Loizides and B. Schmidt},
  organization  = {IOS Press}
}

@Article{         Lafreniere_2007,
  title         = {A New Algorithm for Point-Spread Function Subtraction in
                  High-Contrast Imaging: A Demonstration with Angular
                  Differential Imaging},
  author        = {David Lafrenière and Christian Marois and René Doyon and
                  Daniel Nadeau and Étienne Artigau},
  year          = {2007},
  journal       = {The Astrophysical Journal},
  volume        = {660},
  number        = {1},
  pages         = {770--780},
  doi           = {10.1086/513180},
  eprint        = {astro-ph/0702697},
  eprinttype    = {arXiv}
}

@Article{         Manjon_2015,
  title         = {{MRI} noise estimation and denoising using non-local
                  {PCA}},
  author        = {José V. Manjón and Pierrick Coupé and Antonio Buades},
  year          = {2015},
  month         = {5},
  journal       = {Medical Image Analysis},
  volume        = {22},
  number        = {1},
  pages         = {35--47},
  doi           = {10.1016/j.media.2015.01.004}
}

@Article{         Marois_2006,
  title         = {Angular Differential Imaging: A Powerful High-Contrast
                  Imaging Technique},
  author        = {Christian Marois and David Lafrenière and René Doyon and
                  Bruce Macintosh and Daniel Nadeau},
  year          = {2006},
  journal       = {The Astrophysical Journal},
  volume        = {641},
  number        = {1},
  pages         = {556--564},
  doi           = {10.1086/500401},
  eprint        = {astro-ph/0512335},
  eprinttype    = {arXiv}
}

@Article{         Matplotlib_2007,
  title         = {Matplotlib: A 2D graphics environment},
  author        = {Hunter, J. D.},
  year          = {2007},
  journal       = {Computing in Science \& Engineering},
  volume        = {9},
  number        = {3},
  pages         = {90--95},
  doi           = {10.1109/MCSE.2007.55}
}

@Article{         Mawet_2014,
  title         = {Fundamental Limitations Of High Contrast Imaging Set By
                  Small Sample Statistics},
  author        = {D. Mawet and J. Milli and Z. Wahhaj and others},
  year          = {2014},
  journal       = {The Astrophysical Journal},
  volume        = {792},
  number        = {2},
  pages         = {97},
  doi           = {10.1088/0004-637x/792/2/97},
  eprint        = {1407.2247},
  eprinttype    = {arXiv}
}

@Article{         Numpy_2020,
  title         = {Array programming with {NumPy}},
  author        = {Charles R. Harris and K. Jarrod Millman and Stéfan J. van
                  der Walt and others},
  year          = {2020},
  month         = {9},
  journal       = {Nature},
  volume        = {585},
  number        = {7825},
  pages         = {357--362},
  doi           = {10.1038/s41586-020-2649-2}
}

@InProceedings{   Pandas_2010,
  title         = {Data Structures for Statistical Computing in Python},
  author        = {Wes McKinney},
  year          = {2010},
  doi           = {10.25080/majora-92bf1922-00a},
  booktitle     = {Proceedings of the 9th Python in Science Conference},
  publisher     = {{SciPy}}
}

@Software{        Pandas_2020,
  title         = {pandas-dev/pandas: Pandas 1.1.2},
  author        = {Reback, Jeff and McKinney, Wes and {Jbrockmendel} and
                  others},
  year          = {2020},
  doi           = {10.5281/zenodo.3509134},
  publisher     = {Zenodo},
  copyright     = {Open Access}
}

@Article{         Perrin_2003,
  title         = {The Structure of High Strehl Ratio Point-Spread
                  Functions},
  author        = {Marshall D. Perrin and Anand Sivaramakrishnan and Russell
                  B. Makidon and Ben R. Oppenheimer and James R. Graham},
  year          = {2003},
  journal       = {The Astrophysical Journal},
  volume        = {596},
  number        = {1},
  pages         = {702--712},
  doi           = {10.1086/377689},
  eprint        = {astro-ph/0306468},
  eprinttype    = {arXiv}
}

@Software{        Photutils_2020,
  title         = {astropy/photutils: 1.0.1},
  author        = {Bradley, Larry and Sipőcz, Brigitta and Robitaille,
                  Thomas and others},
  year          = {2020},
  doi           = {10.5281/zenodo.596036},
  publisher     = {Zenodo},
  copyright     = {Open Access}
}

@Article{         Quanz_2010,
  title         = {{First results from very large telescope NACO apodizing
                  phase plate: 4 µm images of the exoplanet $\upbeta$
                  Pictoris b}},
  author        = {Sascha P. Quanz and Michael R. Meyer and Matthew A.
                  Kenworthy and others},
  year          = {2010},
  month         = {9},
  journal       = {The Astrophysical Journal},
  volume        = {722},
  number        = {1},
  pages         = {L49--L53},
  doi           = {10.1088/2041-8205/722/1/l49},
  eprint        = {1009.0538},
  eprinttype    = {arXiv}
}

@Article{         Quanz_2015,
  title         = {{Direct detection of exoplanets in the 3--10\,$\upmu$m
                  range with \mbox{E-ELT}/METIS}},
  author        = {{Quanz}, Sascha P. and {Crossfield}, Ian and {Meyer},
                  Michael R. and {Schmalzl}, Eva and {Held}, Jenny},
  year          = {2015},
  month         = {4},
  journal       = {International Journal of Astrobiology},
  volume        = {14},
  number        = {2},
  pages         = {279--289},
  doi           = {10.1017/S1473550414000135},
  eprint        = {1404.0831},
  eprinttype    = {arXiv}
}

@Article{         Ribak_2008,
  title         = {Fainter and closer: finding planets by symmetry breaking},
  author        = {Erez N. Ribak and Szymon Gladysz},
  year          = {2008},
  month         = sep,
  journal       = {Optics Express},
  volume        = {16},
  number        = {20},
  pages         = {15553},
  doi           = {10.1364/oe.16.015553},
  eprint        = {0809.1825},
  eprinttype    = {arXiv}
}

@Article{         Ruffio_2017,
  title         = {Improving and Assessing Planet Sensitivity of the {GPI}
                  Exoplanet Survey with a Forward Model Matched Filter},
  author        = {Jean-Baptiste Ruffio and Bruce Macintosh and Jason J. Wang
                  and Laurent Pueyo and Eric L. Nielsen and others},
  year          = {2017},
  journal       = {The Astrophysical Journal},
  volume        = {842},
  number        = {1},
  pages         = {14},
  doi           = {1705.05477},
  eprint        = {0809.1825},
  eprinttype    = {arXiv}
}

@Article{         Samland_2020,
  title         = {{TRAP: A temporal systematics model for improved direct
                  detection of exoplanets at small angular separations}},
  author        = {M. Samland and J. Bouwman and D. W. Hogg and W. Brandner
                  and T. Henning and M. Janson},
  year          = {2020},
  eprint        = {2011.12311},
  eprinttype    = {arXiv}
}

@Article{         Schoelkopf_2016,
  title         = {Modeling confounding by half-sibling regression},
  author        = {Bernhard Schölkopf and David W. Hogg and Dun Wang and
                  Daniel Foreman-Mackey and Dominik Janzing and Carl-Johann
                  Simon-Gabriel and Jonas Peters},
  year          = {2016},
  month         = {7},
  journal       = {Proceedings of the National Academy of Sciences},
  volume        = {113},
  number        = {27},
  pages         = {7391--7398},
  doi           = {10.1073/pnas.1511656113}
}

@Article{         Scipy_2020,
  title         = {{{SciPy} 1.0: Fundamental Algorithms for Scientific
                  Computing in Python}},
  author        = {Virtanen, Pauli and Gommers, Ralf and Oliphant, Travis E.
                  and others},
  year          = {2020},
  journal       = {Nature Methods},
  volume        = {17},
  pages         = {261--272},
  doi           = {10.1038/s41592-019-0686-2}
}

@Software{        Seaborn_2020,
  title         = {mwaskom/seaborn: v0.11.0 (September 2020)},
  author        = {Waskom, Michael and Botvinnik, Olga and Gelbart, Maoz and
                  others},
  year          = {2020},
  doi           = {10.5281/zenodo.592845},
  publisher     = {Zenodo},
  copyright     = {Open Access}
}

@Article{         Sivaramakrishnan_2002,
  title         = {Speckle Decorrelation and Dynamic Range in Speckle
                  Noise-limited Imaging},
  author        = {Anand Sivaramakrishnan and James P. Lloyd and Philip E.
                  Hodge and Bruce A. Macintosh},
  year          = {2002},
  month         = {10},
  journal       = {The Astrophysical Journal},
  volume        = {581},
  number        = {1},
  pages         = {L59--L62},
  doi           = {10.1086/345826}
}

@Article{         Sklearn_2011,
  title         = {Scikit-learn: Machine Learning in {Python}},
  author        = {Pedregosa, F. and Varoquaux, G. and Gramfort, A. and
                  others},
  year          = {2011},
  journal       = {Journal of Machine Learning Research},
  volume        = {12},
  pages         = {2825--2830}
}

@Article{         Soummer_2012,
  title         = {Detection And Characterization of Exoplanets And Disks
                  Using Projections On Karhunen-Loève Eigenimages},
  author        = {Rémi Soummer and Laurent Pueyo and James Larkin},
  year          = {2012},
  journal       = {The Astrophysical Journal},
  volume        = {755},
  number        = {2},
  pages         = {L28},
  doi           = {10.1088/2041-8205/755/2/l28},
  eprint        = {1207.4197},
  eprinttype    = {arXiv}
}

@Article{         Stolker_2019,
  title         = {{PynPoint}: a modular pipeline architecture for processing
                  and analysis of high-contrast imaging data},
  author        = {T. Stolker and M. J. Bonse and S. P. Quanz and A. Amara
                  and G. Cugno and A. J. Bohn and A. Boehle},
  year          = {2019},
  journal       = {Astronomy {\&} Astrophysics},
  volume        = {621},
  pages         = {A59},
  doi           = {10.1051/0004-6361/201834136},
  eprint        = {1811.03336},
  eprinttype    = {arXiv}
}

@Article{         Wang_2017,
  title         = {A pixel-level model for event discovery in time-domain
                  imaging},
  author        = {Dun Wang and David W. Hogg and Daniel Foreman-Mackey and
                  Bernhard Schölkopf},
  year          = {2017},
  eprint        = {1710.02428},
  eprinttype    = {arXiv}
}
